\documentclass[sigconf]{acmart}
\usepackage{graphicx}  
\usepackage{latexsym}
\usepackage{CJKutf8}
\usepackage{multirow}
\usepackage{booktabs}
\usepackage{enumitem}
\usepackage{amsmath}
\usepackage{amssymb}
\usepackage{multirow}
\usepackage{array}
\usepackage{xcolor}
\usepackage{subfig}
\usepackage[boxruled,noend]{algorithm2e}
\usepackage{caption}

\copyrightyear{2020}
\acmYear{2020}
\setcopyright{acmlicensed}\acmConference[CIKM '20]{Proceedings of the 29th ACM International Conference on Information and Knowledge Management}{October 19--23, 2020}{Virtual Event, Ireland}
\acmBooktitle{Proceedings of the 29th ACM International Conference on Information and Knowledge Management (CIKM '20), October 19--23, 2020, Virtual Event, Ireland}
\acmPrice{15.00}
\acmDOI{10.1145/3340531.3411884}
\acmISBN{978-1-4503-6859-9/20/10}

\newcolumntype{H}{>{\setbox0=\hbox\bgroup}c<{\egroup}@{}}

\begin{document}

\title{Attacking Recommender Systems with Augmented User Profiles}

\author{Chen Lin}
\authornote{Chen Lin is supported by the National Natural Science Foundation of China (no.61972328).}
\affiliation{Xiamen University}
\email{chenlin@xmu.edu.cn}

\author{Si Chen}
\affiliation{Xiamen University}
\email{sichen@stu.xmu.edu.cn}

\author{Hui Li}
\authornote{Hui Li is the corresponding author.}
\affiliation{Xiamen University}
\email{hui@xmu.edu.cn}

\author{Yanghua Xiao}
\affiliation{Fudan University}
\email{shawyh@fudan.edu.cn}

\author{Lianyun Li}
\affiliation{Xiamen University}
\email{lilianyun@stu.xmu.edu.cn}

\author{Qian Yang}
\affiliation{Xiamen University}
\email{yangqian@stu.xmu.edu.cn}

\begin{abstract}
Recommendation Systems (RS) have become an essential part of many online services. Due to its pivotal role in guiding customers towards purchasing, there is a natural motivation for unscrupulous parties to spoof RS for profits. In this paper, we study the shilling attack: a subsistent and profitable attack where an adversarial party injects a number of user profiles to promote or demote a target item. Conventional shilling attack models are based on simple heuristics that can be easily detected, or directly adopt adversarial attack methods without a special design for RS. Moreover, the study on the attack impact on deep learning based RS is missing in the literature, making the effects of shilling attack against real RS doubtful. We present a novel Augmented Shilling Attack framework (AUSH) and implement it with the idea of Generative Adversarial Network. AUSH is capable of tailoring attacks against RS according to budget and complex attack goals, such as targeting a specific user group. We experimentally show that the attack impact of AUSH is noticeable on a wide range of RS including both classic and modern deep learning based RS, while it is virtually undetectable by the state-of-the-art attack detection model.
\end{abstract}

% <concept_id>10002951.10003317.10003347.10003350</concept_id>
% <concept_desc>Information systems~Recommender systems</concept_desc>

% \ccsdesc[500]{Information systems~Recommender systems}


% \keywords{Shilling Attack, Recommender Systems, Generative Adversarial Network}

\maketitle

\section{Introduction}
\label{sec:intro}

The history of \textbf{R}ecommender \textbf{S}ystems (RS) can be traced back to the beginning
of e-commerce~\citep{Aggarwal16}. The ability of RS to assist users in finding
the desirable targets makes it an important tool for alleviating information
overload problem. As a result, RS has been prevalently deployed in industries (e.g.,
Amazon, Facebook and Netflix~\citep{Aggarwal16}). Not only is RS beneficial
to customers, but also RS helps retail companies and producers promote their
products and increase sales. Consequently, there is a strong intention for unscrupulous 
parties to attack RS in order to maximize their
malicious objectives. 

Due to RS's pivotal role in e-commerce, much effort has been devoted to studying how to spoof RS in order to give insights into the defense against malicious attacks. 
Various attacks, such as
unorganized malicious attack (i.e., several attackers individually attack RS
without an organizer)~\citep{Pang0TZ18} and sybil attack (i.e., illegally infer a user's
preference)~\citep{CalandrinoKNFS11}, have been studied. This paper
focuses on a subsistent and profitable attack, i.e., \emph{shilling
attack}, where an adversarial party produces a number of user profiles using some
%strategies to promote (a.k.a. push attack) or demote (a.k.a. nuke attack) an
strategies to promote or demote an
item~\citep{GunesKBP14} in order to have their own products recommended more often
than those of their competitors. Shilling attack is also called data
poisoning~\citep{LiWSV16} or profile injection attack~\citep{BurkeMBW05} in
the literature. Researchers have successfully performed shilling attacks
against real-world RS such as YouTube, Google Search, Amazon and Yelp in
experiments~\citep{XingMDSFL13,YangGC17}. Large companies like Sony,
Amazon and eBay have reported that they suffered from such attacks in
practice~\citep{LamR04}. 

%\begin{figure}[htbp]
%\begin{center}
%\includegraphics[width=0.9\columnwidth]{example.eps}
%\caption{An illustrative example of shilling attack}
%\label{default}
%\end{center}
%\end{figure}

Shilling attack is the specific application of \emph{adversarial
attack}~\citep{abs-1810-00069,YuanHZL19} in the domain of recommender systems.
Adversarial attack uses crafted adversarial examples to mislead machine
learning models.  A tremendous amount of work in adversarial attack is against image classification~\citep{SuVS19}, or text classification~\citep{AlzantotSEHSC18}. However, they cannot be directly employed in shilling attack at full power, due to the following challenges: 
\begin{enumerate}[label=(\arabic*),leftmargin=15pt,topsep=2pt]
\item \textbf{Data correlation in RS:} RS relies on capturing the correlations between users and items for recommendations and such relations enhance the robustness of RS. The recommendation targeting at a specific user is typically made based on the information from multiple user-item pairs (i.e., collaborative filtering~\citep{Aggarwal16}) instead of a single data sample. Therefore, manipulating the recommendation for one user requires to inject many related user-item pairs, which may affect the recommendation results for other non-targeting users in RS and make the attack easy to be detected. This is different compared to attacking many other learning tasks where manipulating one data sample may achieve the desirable attack goal and adversarial attacks can be directly deployed (e.g., one-pixel attack~\cite{SuVS19}).  %  for changing the classification of an image
\item \textbf{No prior knowledge of RS:} A prevalent strategy of adversarial attack is to utilize the information of gradient decent in machine learning models to search undetectable perturbations and then combine perturbations with normal representation vectors to affect the learning system~\citep{abs-1810-00069}. As a comparison, in shilling attack, though the data (e.g., rating matrix) of RS is generally available to all users (i.e., a user can see all other users' ratings) and thus exposed to attackers~\cite{SandvigMB08,GunesKBP14,LamR04}, the recommendation model is typically a \textit{black box}. Thus, it is required that the attack must be effective against a wide range of recommendation models. 
\item \textbf{The balance of different complex attack goals:} Instead of only promoting or demoting an item to the general audience, there are usually multiple goals that the attacker desires to achieve. However, incorporating multiple attack goals together may degrade the attack performance of individual attack goal or make the attack detectable. Consequently, special designs are required to balance and achieve multiple attack goals simultaneously, while keeping the attack undetectable.
\end{enumerate}

Due to the aforementioned challenges, only a few recent works~\cite{Christakopoulou2018Adversarial,Christakopoulou19} consider directly adopting the idea of adversarial attacks for shilling attack, and they do not show satisfactory attack effects on a wide range of RS as illustrated later in our experiments.
In addition to these methods, most existing shilling attack methods create injection profiles based on some global statistics, e.g., average rating value~\citep{GunesKBP14,LamR04} and rating variance~\citep{SandvigMB08} for each item. For instance, average attack assigns the
highest rating to the target item to be promoted and an average rating to a
set of randomly sampled items~\citep{LamR04}. Although all these existing methods, including both adversarial based and simple heuristic based approaches, were proved to be effective in some cases, they still suffer from the following limitations: 
\begin{enumerate}[label=(\arabic*),leftmargin=15pt,topsep=2pt]
\item \textbf{Easy to detect}: Generated user profiles lack personalization (i.e., different user behavior pattern), thus the injected profiles can be easily detected, even by some simple heuristics (more details described in Sec.~\ref{sec:detection}). 
\item \textbf{Narrow range of target models}: Depending on how the statistics are computed, conventional shilling attacks are shown to be effective only on certain traditional collaborative filtering (CF) approaches. For example, average, bandwagon and random attacks are more effective against user-based KNN, but do not work well against item-based KNN~\citep{MobasherBBW07}.
Moreover, their influence on deep learning based RS, which has attracted considerable interest and been deployed in real applications~\citep{ZhangYST19}, has not been studied. 
In fact, as global statistics can not capture high-level associations among items or users, the
actual effect of the existing attack approaches on modern RS is doubtable (more details described in Sec.~\ref{sec:attack}). 
\item \textbf{Inflexibility}: It is difficult to tailor the attack for specific goals which attackers desire to achieve after the attack, e.g., to exert adverse effects on items
from the competitors.
\end{enumerate}

To address the above problems, a natural intuition to enhance the
attack is to ``augment" the templates, which are selected from existing real
user profiles and are used to generate injected profiles. This way, the
injected fake user profiles are diversified and it becomes difficult to
distinguish them from real users. Based on this intuition, we present a novel
\textbf{Au}gmented \textbf{Sh}illing Attack (AUSH) and implement it with the
idea of Generative Adversarial Network (GAN)~\citep{GoodfellowPMXWOCB14}.
Specifically, the generator acts like an ``attacker'' and generates fake user
profiles by augmenting the ``template'' of existing real user profiles.  The
deep neural network based generator can capture complex user-item associations
better than existing attack methods using simple heuristics. Thus, it works
well on modern RS which commonly deploys deep neural networks. Moreover, the
generator is able to achieve secondary attack goals by incorporating a
shilling loss.  On the other hand, the discriminator module performs like a
``defender''. It distinguishes fake user profiles from real user profiles and
provides guidance to train the generator to generate undetectable fake user
profiles. Each of the generator and the discriminator strikes to enhance
itself to beat the other one at every round of the minimax competition. 
It is worthy noting that, as we have explained, deploying the idea of adversarial attack in shilling attack is not a trivial task, and \emph{directly applying the adversarial attack method (i.e., using a general GAN) in shilling attacks
without our designs to tailor it for the attack will not provide satisfactory
results as shown in our experiments}.

Our contributions can be summarized by three merits of AUSH. We show that AUSH resembles to the traditional \emph{segment attack} and
\emph{bandwagon attack}~\citep{GunesKBP14}, yet more \emph{powerful},
\emph{undetectable} and \emph{flexible} than conventional shilling attack
methods:
\begin{enumerate}[label=(\arabic*),leftmargin=15pt,topsep=2pt]
\item AUSH is powerful on a wide range of recommendation models including
both traditional CF methods and modern deep learning based approaches, while
\emph{the prior knowledge of AUSH does not exceed what the conventional shilling
attack approaches require to know}. 
% Note that attacking modern deep neural
% network recommendation algorithms has rarely been studied before. 
\item Furthermore, AUSH is virtually undetectable by the state-of-the-art attack detection method
as shown in our experiments. 
\item Finally, AUSH contains more than a general GAN as it includes a reconstruction loss and a shilling loss which tailor AUSH for attacking RS and endows the AUSH with the ability of achieving secondary attack goals (e.g., promote
items for a group of users who have shown preferences over a predefined set of
competitors, or target on long-tail items). 
\end{enumerate}
We conduct comprehensive experiments to verify the above merits of AUSH and its attack power against both classic and modern deep learning based recommendation algorithms. Note that attacking modern deep neural network recommendation algorithms has rarely been studied in the literature. 

% In summary, our contributions in this paper are three-fold: 
% \begin{enumerate}[label=(\arabic*),leftmargin=20pt]
% \item We present a novel framework to generate virtually undetectable fake user profiles (according to state-of-the-art fake detection models) using the idea of GAN to fulfill shilling attack goals on RS. 
% We show that conventional attack models such as segment attack and bandwagon attack can be expressed under this framework. 

% \item AUSH contains more than a general GAN. It includes a reconstruction loss and a shilling loss which tailors AUSH for attacking RS and endows the AUSH with ability of achieving multiple attack goals. We further present various sampling strategies to provide more options for shilling attacks. 

% \item We conduct comprehensive experiments to show that AUSH is more powerful (in terms of prediction shift and hit ratios) against a wide range of classic and modern recommendation algorithms, and undetectable to attack detection methods, compared with conventional attack strategies. Note that attacking modern deep neural network recommendation algorithms has rarely been studied. 
% \end{enumerate}

The rest of the paper is organized as follows: Sec.~\ref{sec:related}
illustrate the related work. Sec.~\ref{sec:model} demonstrates the design of
AUSH and Sec~\ref{sec:implementation} gives one possible implementation of
AUSH. In Sec.~\ref{sec:experiment}, we compare AUSH with other
state-of-the-art shilling attacks methods and verify its effectiveness.
Sec.~\ref{sec:con} concludes our work.

\section{Related Work}
\label{sec:related} 

% We briefly survey four lines of research related to our work.

\subsection{Recommender Systems (RS)}
Traditional RS typically relies on collaborative filtering methods (CF),
especially matrix factorization (MF) methods~\citep{LiCYM17}. MF models user preferences and
item properties by factorizing the user-item interaction matrix into two
low-dimensional latent matrices. Recently, numerous deep learning techniques (e.g., MLP~\citep{HeLZNHC17}, CNNs~\citep{TuanP17}, RNNs~\citep{SunW018},
GNNs~\citep{FanZHSHML19}, Autoencoder~\citep{Sedhain2015AutoRec}, and the Attention
Mechanism~\citep{LiYZJ20}) have been introduced into RS. Compared to
traditional RS, deep learning based RS is able to model the nonlinearity of data correlations and learn the underlying complex feature representations~\citep{ZhangYST19}. Consequently, deep
learning based RS has outperformed traditional RS in general.

\subsection{Adversarial Attacks} 
Machine learning has played a vital role in a broad spectrum of applications and helped solve many difficult problems for us. However, security of machine learning systems are vulnerable to crafted adversarial examples~\citep{abs-1810-00069}, which may be imperceptible to the human eye, but can lead the model to misclassify the output. Adversaries may leverage such vulnerabilities to compromise a learning system where they have high incentives and such attacks are called as \emph{adversarial attacks}. Adversarial attack has show its ability to manipulate the outputs of many text and image based learning systems~\citep{abs-1902-07285,Zhang20,abs-1810-00069,YuanHZL19}.

Adversarial examples in conventional machine learning models have been
discussed since decades ago~\citep{YuanHZL19}. \citet{DalviDMSV04} find
manipulating input data may affect the prediction results of classification
algorithms. \citet{BiggioCMNSLGR13} design a gradient-based approach to
generate adversarial examples against SVM. \citet{BarrenoNSJT06,BarrenoNJT10}
formally investigate the security of conventional machine learning methods
under adversarial attacks. \citet{RoliBF13} discuss several defense strategies
against adversarial attacks to improve the security of machine learning
algorithms. In addition to conventional machine learning, recent studies have reported that deep learning techniques are also vulnerable to adversarial attacks~\citep{abs-1810-00069,YuanHZL19}.

Though we have witness a great success of adversarial attacks against many learning systems, existing adversarial attacks cannot be directly adopted for the shilling attack task as explained in Sec.~\ref{sec:intro}.

\vspace{-5pt}
\subsection{Generative Adversarial Network}
Generative Adversarial Network (GAN)~\citep{GoodfellowPMXWOCB14} has recently
attracted great attention for its potential to learn real data distribution
and generate text~\citep{YuZWY17}, images~\citep{LiuT16}, recommendations~\citep{WangNL2019} and many other types of
data~\citep{HongHYY19}. GAN performs adversarial learning
between the generator and the discriminator. The generator and the
discriminator can be implemented with any form of differentiable system that
maps data from one space to the other. The generator
tries to capture the real data distribution and generates real-like data,
while the discriminator is responsible for discriminating the data generated
by the generator and the real data. GAN plays a minimax game and the
optimization terminates at a saddle point that is a minimum with respect to
the generator and a maximum with respect to the discriminator (i.e., Nash
equilibrium). 

As GAN overcomes the limitations of previous generative
models~\citep{HongHYY19}, it has been successfully applied in many
applications and there is a surge of works studying how to improve
GAN~\citep{HongHYY19}. Follow-up works include
DCGAN~\citep{RadfordMC15} which adopts the CNN architecture in GAN and
Wasserstein GAN~\citep{ArjovskyCB17} which leverages Earth Mover distance.
There also exists a direction of GAN research which utilizes GAN to generate
adversarial examples. For instance, \citep{ZhaoDS18} propose to search the
representation space of input data instead of input data itself under the
setting of GAN in order to generate more natural adversarial examples.
\citep{XiaoLZHLS18} design AdvGAN which can attack black-box models by
training a distilled model. 

\vspace{-5pt}
\subsection{Shilling Attacks against RS} 
\citet{OMahonyHS05,OMahonyHKS04} firstly study the robustness of
user-based CF method for rating prediction by injecting some faked users. They
also provide a theoretical analysis of the attack by viewing injected ratings
as noises. \citet{LamR04,BurkeMBW05,Burke2005Limited,MobasherBBW07} 
further study the influence of some low-knowledge attack approaches to promote an item (e.g., %push attack approaches (e.g., 
random, average, bandwagon attack and segment attack) and to demote an item %nuke attack methods
(e.g., love/hate attack and reverse bandwagon attack) on CF methods
for both rating prediction and top-$K$ recommendation. They observe that CF
methods are vulnerable to such attacks.
Assuming more knowledge and cost, \citet{WilsonS13,SeminarioW14b} design the
power user/item attack models which leverage most influential users/items to
shill RS, \citet{FangYGL18} study how to shill a graph based CF models, and \citet{LiWSV16} present near-optimal data poisoning attacks for factorization-based CF.
\citet{XingMDSFL13,YangGC17} conduct experiments on attacking real-world
RS (e.g., YouTube, Google Search, Amazon and Yelp) and show that manipulating
RS is possible in practice. 

Inspired by the success of GAN, a few works turn to leverage GAN for shilling attack task~\cite{Christakopoulou2018Adversarial,Christakopoulou19}. 
However, directly adopting existing GAN methods for generating adversarial examples, without special designs (like AUSH) to tailor them for RS, will not provide satisfactory results in shilling attacks as shown in our experiments.
\citet{Christakopoulou2018Adversarial,Christakopoulou19} employ DCGAN~\citep{RadfordMC15} to generate faked user
profiles used in shilling attacks. They formulate this procedure as a repeated
general-sum game between RS and adversarial fake user generator. Compared to
their work, AUSH is more specially tailored for RS instead of directly using
adversarial attacks (i.e., the general GAN) against machine learning models.
We consider more realistic factors (e.g., users in the segment, attack cost
and undetectability) when attacking RS, which descend from previous study on
attacking traditional CF models. 

Note that the study on the impact of shilling attacks against deep learning based RS is limited, although there is a tremendous amount of work on attack and defense of traditional RS. Therefore, we also include an analysis of attacking deep learning based RS in the Sec.~\ref{sec:experiment} of this paper. 
% \hui{Ignore \citep{Song20}?}

\section{Augmented Shilling Attack}
\label{sec:model}
In this section, we introduce our proposed attack framework: Augmented Shilling Attack (AUSH).

\begin{figure*}[t]
\begin{center}
\includegraphics[width=1\textwidth]{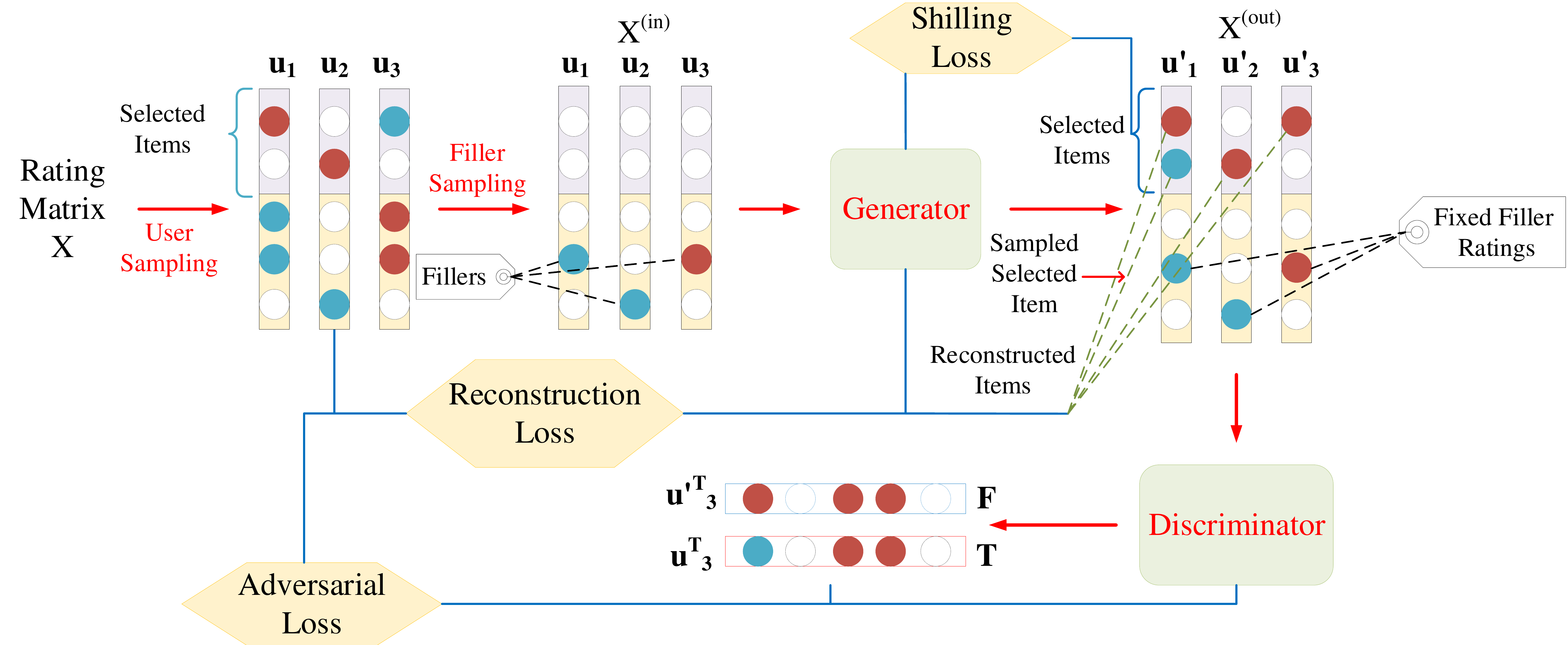}
\caption{Pipeline of AUSH. We use binary ratings for illustration, though AUSH can handle a five-point scale. Red and blue indicate a high rating and a low rating, respectively.}
\label{fig:framework}
\end{center}
\vspace{-5pt}
\end{figure*}

\subsection{Terminology}
We follow the terminology used in the literature~\citep{GunesKBP14} and divide
the items in a fake user profile into one \emph{target item} (i.e., the
attacker wants to assign it a malicious rating), a number of \emph{filler
items} (i.e., a group of randomly sampled items which have been rated by the
real user and will be used to obstruct detection of the attack), a number
of \emph{selected items} (i.e., a group of human-selected items for special
treatment to form the characteristics of the attack), and \emph{unrated items}
(i.e., the rest of the items in the RS). Selected items are the same across
all fake user profiles, while each fake user profile has its own filler items.

\subsection{Attack Budget and Goal}
\label{sec:goal}
Attacking RS is costly. 
As such, in designing a practical attack model against RS, we have to take into account the following attack budget and goal:

\begin{itemize}[leftmargin=10pt]
\setlength{\itemsep}{3pt}
\item \textbf{Attack budget}: we consider two factors
	\begin{itemize}
	\item \textbf{Attack size} is the number of fake user profiles
	\item \textbf{Profile size} is the number of non-zero ratings. The larger the attack size / profile size is, the more effective and expensive the attack could be.
	\end{itemize}
\item \textbf{Attack goal}: the goal an adversarial party wants to achieve could be complex and we mainly consider the following aspects
	\begin{itemize}
		\item \textbf{Attack type} indicates whether it is a push attack (i.e., assign a maximal rating on target item to promote it) or a nuke attack (i.e., assign a minimal rating on target item to demote it). Since the two types are similar and can be exchanged (i.e., change a maximal rating to a minimal rating), we consider push attacks in the sequel for simplicity. 
		\item \textbf{Target user group} is the group of users that an attack aims at. 
		\item \textbf{Ancillary effects} (e.g., demoting competitors, bias the ratings of a special user groups on selected items) are also desired in the attack. Such intentions will manifest in choosing selected items.
	\end{itemize}
\end{itemize}

\vspace{-10pt}
\subsection{Overview of AUSH}

Conventional attack models make up a fake user profile from scratch. 
On the contrary, our intuition is to use an existing real user profile as a ``template'' and \textbf{au}gment it to generate the fake user profile for \textbf{sh}illing (AUSH). The template knowledge is accessible in practice and do not exceed the requirements of recent sophisticated attack methods~\citep{LiWSV16,FangYGL18}, as we will show later. 
The benefits are two-fold.
Firstly, the generated profile is indistinguishable as it is built upon real user behavior patterns. 
Moreover, it retains the preference diversity of the community. Unlike random or average attack, where fake users do not show specific tastes, our strategy can generate fake users who have a special taste on niche items.  

Inspired by the success of adversarial learning in image generation~\citep{GoodfellowPMXWOCB14}, we employ a Generative Adversarial Network framework for AUSH to make the attack even more undetectable. 
Fig.~\ref{fig:framework} gives an overview of our pipeline, which consists of the following parts:
\begin{enumerate}[label=(\arabic*),leftmargin=15pt]
\item \textbf{Sampling (``template" selection)} contains two steps. In the first step, a batch of real users are chosen as ``templates''. In the second step, filler items of each ``template'' are sampled from the rated items of the corresponding ``template'' user. 
\item \textbf{Generator (``patch" generation)} patches each ``template" by
adding ratings on selected items to generate one fake profile.
Generator takes as input the sampled user-item rating sub-matrix (i.e., ``templates'') and captures the latent association between items and users. 
To better learn behavior patterns of the real user (i.e., the ``template'' user) including positive and negative preference on selected items, AUSH attempts to recover each ``template'' user's observed ratings on selected items and samples of unobserved selected items (i.e., to recover the rating ``$0$") via a \emph{reconstruction loss}. The output of generator is a set of fake user profiles, which contain ratings on selected items. We can harness a \emph{shilling loss} to optimize secondary attack effects, including but not limited to demoting the competitors, targeting on special user groups, etc. 
\item \textbf{Discriminator} is fed with the output of the generator. It attempts to accurately classify real user profiles and fake user profiles. The \emph{adversarial loss} is optimized to boost the performance of discriminator.
\end{enumerate}

The design of AUSH is general and there are various possible implementations for the generator and the discriminator. We provide one implementation in Sec.~\ref{sec:implementation}.

\setlength{\textfloatsep}{5pt} % space after algorithm
\begin{algorithm}[t]
\caption{Training procedure for AUSH}\label{alg:training}
\KwIn{rating matrix $\mathbf{X}$}
\KwOut{parameter set $\mathbf{\mathbf{\theta}}$ for generator $G$ and parameter set $\mathbf{\phi}$ for discriminator $D$}
\For{number of training epochs}
{
	\For{$k_1$ steps}
	{
		uniformly sample a minibatch of users $\mathcal{U}'$\;
			\ForEach{ $u'\in \mathcal{U'}$}
			{
				sample $F$ items to construct $\mathbf{x}^{(in)}_{u'}$\;
			}
		generate a minibatch of fake user profiles $\{\mathbf{x}^{(out)}_{u'}=G(\mathbf{x}^{(in)}_{u'})| u'\in \mathcal{U'}\}$\;
		optimize $\mathbf{\phi}$ to $\max H(G,D)$ with $\mathbf{\theta}$ fixed\;
	} %only one step 

	\For{$k_2$ steps}
	{
		uniformly sample a minibatch of user rating vectors $\{\mathbf{x}_{u}\}$\;
		\ForEach{ $u'\in \mathcal{U'}$}
		{
			sample $F$ items to construct $\mathbf{x}^{(in)}_{u'}$\;
		}
		generate a minibatch of fake user profiles $\{\mathbf{x}^{(out)}_{u'}=G(\mathbf{x}^{(in)}_{u'})| u'\in \mathcal{U'}\}$\;
		optimize $\mathbf{\theta}$ to $\min \mathcal{L}_{AUSH}$ with $\mathbf{\phi}$ fixed\;
	}
}

\end{algorithm}

\subsection{Relation to Segment/Bandwagon Attack}\label{sec:relation}

Segment attack injects user profiles, each of which comprises maximal ratings on selected items and minimal ratings on filler items. For \emph{in-segment users} (defined as users who like selected items), segment attack is one of the few attack models that work effectively on item-based CF recommendation models. 
The design of segment attack ensures that similarity between users in the segment and injected profiles appears high and target item becomes more likely to be recommended. 

Another commonly adopted attack model is bandwagon attack. 
In bandwagon attack, the most popular items are regarded as selected items and are assigned with highest ratings. The filler items are randomly chosen and randomly rated. 
It associates the target item with popular items, so that the inserted profiles will have a high probability of being similar to many users~\citep{Burke2005Limited}.

We see that segment attack and bandwagon attack can be expressed under our
framework. If we fix ratings on the fillers and
selected items to be minimal rating and maximal rating respectively, then AUSH is degraded to segment attack.  If we sample
frequently rated items as selected items, then AUSH is degraded to bandwagon attack.
Due to the architectural resemblance, AUSH is empowered by the capabilities of both segment attack and bandwagon attack.
Moreover,
AUSH improves over bandwagon attack by
allowing the selected item to be tuned according to the rating values of
fillers, making the injected profile more natural and indistinguishable. 
It also advances segment attack by revealing real patterns of filler ratings.
In addition to the aforementioned advantages, AUSH is more flexible and able to
achieve multiple goals (i.e., \emph{Attack Goal} in Sec.~\ref{sec:goal}) in
a single attack.

\section{Implementation}
\label{sec:implementation}

We use $\mathbf{X}\in \mathcal{R}^{|\mathcal{V}|\times |\mathcal{U}|}$ to denote the rating matrix in RS, where $\mathcal{U}$ is the set of real users and $\mathcal{V}$ is the item universe. 
$\mathcal{V}_u=\{v\in \mathcal{V}: x_{v,u}\neq 0\}$ indicates the set of items that have been rated by $u$.
Similarly,  $\mathcal{U}_v=\{u\in \mathcal{U}: x_{v,u}\neq 0\}$ denotes the set of users that have rated $v$.
Unless otherwise stated, we use lower-case letters for indices, capital letters for scalars, boldface lower-case letters for vectors, boldface capital letters for matrices, calligraphic letters for sets.   
For instance, $A$, $P$, $F$, $\mathcal{U'}$ and $\mathcal{S}$ are attack size, profile size, filler size, set of fake users and set of selected items, respectively. The generator of 
AUSH takes $\mathbf{X}^{(in)}\in \mathcal{R}^{|\mathcal{V}|\times |\mathcal{U'}|}$ as the input and generates the fake user profiles $\mathbf{X}^{(out)}\in \mathcal{R}^{|\mathcal{V}|\times |\mathcal{U'}|}$, where each column has exactly $P$ non-zero entries. As depicted in Alg.~\ref{alg:training}, AUSH comprises of the following components and steps.  

\subsection{Sampling}
\label{sec:sampling}
In this step, AUSH samples a sub-matrix $\mathbf{X}^{(in)}\in \mathcal{R}^{|\mathcal{V}|\times |\mathcal{U'}|}$ (i.e., ``templates'') from $\mathbf{X}$. 
Each ``template'' is sampled randomly from real users who have sufficient ratings. Mathematically, $\forall u'\in\mathcal{U'}, |\mathcal{V}_{u'}|\geq P$. 
In each training epoch of Alg.~\ref{alg:training}, the set $\mathcal{U'}$ is a minibatch of users. 
In test time (i.e., the generated fake profiles are used for attack), we sample exactly $A$ fake user profiles $|\mathcal{U'}|=A$. We adopt different strategies as shown below to sample the filler items for each $u'\in\mathcal{U'}$ and form $\mathbf{x}_{u'}^{(in)}$. For each filler item $v$, $\mathbf{X}_{v,u'}^{(in)}=\mathbf{X}_{v, u'}$. For other items, $\mathbf{X}_{v,u'}^{(in)}=0$. 
\begin{enumerate}[label=(\arabic*),leftmargin=17pt]
\setlength{\itemsep}{2pt}
\setlength{\parskip}{0pt}
\setlength{\parsep}{0pt}
\item \textbf{Random Sample}: randomly sample items from $\mathcal{V}_u$. 
\item \textbf{Sample by Rating}: sample items based on their ratings, i.e., $P\big(\mathbf{X}_{v,u'}^{(in)}\neq 0\big) = \frac{\bar{r}_v }{\sum_{\hat{v}\in \mathcal{V}_u} \bar{r}_{\hat{v}}}$, where $\bar{r}_{v}$ is $v$'s average rating. 
\item \textbf{Sample by Popularity}: items are sampled based on their popularity, i.e., $P\big(\mathbf{X}_{v,u'}^{(in)}\neq 0\big) = \frac{|\mathcal{U}_v|}{\sum_{\hat{v}\in \mathcal{V}_u} |\mathcal{U}_{\hat{v}}| }$. 
\item \textbf{Sample by Similarity}: sample items based on their similarity to the \emph{selected items}, i.e., $P\big(\mathbf{X}_{v,u'}^{(in)}\neq 0\big) = \frac{|\mathcal{U}_v \bigcap \mathcal{U}_{\mathcal{S}}|}{\sum_{\hat{v}\in \mathcal{V}} |\mathcal{U}_{\hat{v}}\bigcap \mathcal{U}_{\mathcal{S}}| }$.
\end{enumerate}

\subsection{Generator}
\label{sec:generator}
The generator aims to ``patch'' the ``templates'' with ratings on selected items in order to form the fake user profiles for attack. 

We employ a \emph{reconstruction loss} (i.e., MSE loss), shown in Eq.~\ref{equ:reconstruction}, to optimize the generator parameters. We will slightly abuse the notation, and define $\mathcal{S}_{u'}^{+}= \mathcal{V}_{u'}\bigcap \mathcal{S}$ as the set of observed ratings of the ``template'' user for user $u'$ on selected items, and $\mathcal{S}_{u'}^{-}= (\mathcal{V}-\mathcal{V}_{u'})\bigcap \mathcal{S}$ as random samples from the set of selected items that the ``template'' user has not rated in the original data. And $T_{u'}$ indicates $\mathcal{S}_{u'}^{+}\bigcup \mathcal{S}_{u'}^{-}$. 
\begin{equation}
\label{equ:reconstruction}
\mathcal{L}_{Recon}=\mathbb{E}_{u'\sim \mathcal{U'}} \sum_{j\in T_{u'}} \big(\mathbf{X}^{(out)}_{j,u'}-\mathbf{X}_{j,u'}\big)^2,\,\,
\mathbf{X}^{(out)}=G\big(\mathbf{X}^{(in)}\big),
\end{equation}
where $G(\cdot)$ indicates the generator which will be defined in Eq.~\ref{eq:generator}.

The reconstruction loss helps to produce ratings on the selected items that are consistent with the real user's preference. Note that we use minibatch for training as shown in Alg.~\ref{alg:training}. Thus we sample $m$ (the percentage) of unobserved selected items for all the users in a minibatch when constructing reconstructed items for these users, instead of independently sampling unobserved selected items for each user.

There is a variety of model structures for optimizing the reconstruction loss, 
we empirically find that towered multilayer perceptron (MLP) combined with the MSE loss on selected items works best. 
% $G:\mathcal{R}^{|\mathcal{V}|} \leftarrow \mathcal{R}^{|\mathcal{V}| } $
Let $N$ be the number of hidden layers, the generator $G$ is a mapping function that operates in a towered manner:
\begin{equation}
\label{eq:generator}
G(\mathbf{x})=f_{out}\bigg(f_N\Big(\cdots f_2\big(f_1(\mathbf{x})\big)\cdots\Big)\bigg).
\end{equation}

In Eq~\ref{eq:generator}, $f_l(\cdot)$ with $l=1,2,\cdots,N$ denotes the mapping function for the $l$-th hidden layer. $f_l(\mathbf{x}) = \sigma(\mathbf{W}_l \mathbf{x}+{\mathbf{b}_l})$, where $\mathbf{W}_l$ and $\mathbf{b}_l$ are learnable weight matrix and bias vector for layer $l$. The activation function $\sigma$ for each layer is sigmoid.
We set the size of layers (i.e., dimensionality of $\mathbf{x}$) as one third of the previous layers. The output layer $f_{out}(\cdot)$ is similar to $f_l(\cdot)$ and its size is the number of selected items.

AUSH can be extended to achieve secondary attack goals by incorporating a \emph{shilling loss}. 
In this work, we consider enhancing the attack effect on in-segment users~\citep{LamR04}. 
That is, we increase the impact of the attack on users who like the selected items before the attack. 
Such an effect is desirable when an adversarial party (i.e., the attacker) is competing with the selected items (from its competitor). The shilling loss we adopt is shown as follows:
% \begin{equation}
% \label{equ:shilling}
% \mathcal{L}_{Shilling} = \mathbb{E}_{u\sim P} \mathbb{E}_{v\in\mathcal{S}} \|Q-G(\mathbf{X'})_{u,v}\|_2,
% \end{equation}
% \begin{small}
% \begin{equation}
% \label{equ:shilling}
% \mathcal{L}_{Shill} = \mathbb{E}_{u'\sim \mathcal{U'}} \mathbb{E}_{\mathbf{s}_{u'} \sim \mathcal{S}} \|\mathbf{q}-G(\mathbf{s}_{u'})\|^2_2,\nonumber
% \end{equation}
% \end{small}
\begin{equation}
\label{equ:shilling}
\mathcal{L}_{Shill} = \mathbb{E}_{u'\sim \mathcal{U'}} \sum_{j\in\mathcal{S}}\big(Q-\mathbf{X}_{j,u'}^{(out)}\big)^2,
\end{equation}
%\mathcal{L}_{Shill} = \mathbb{E}_{u'\sim \mathcal{U'}} \|\mathbf{q}-G(\mathbf{s}_{u'})\|^2_2,\nonumber
where $Q$ is the maximal possible rating in the system.
% , and $G(\mathbf{r}_{u'})_j$ is the ratings of injected user profile on a selected item $j$. 
The shilling loss produces fake user profiles that are more likely to associate with in-segment users. Thus in-segment users, after our attack, prefer to purchase the target item rather than the selected items (from the competitor). Through optimizing shilling loss, AUSH is able to achieve the ancillary effects.
\subsection{Discriminator}
The discriminator $D$ attempts to correctly distinguish fake user profiles from real user profiles, and encourages the generator to produce realistic user profiles. 
We use MLP $D(\mathbf{x}) = \sigma(\mathbf{W} \mathbf{x}+\mathbf{b})$ as our discriminator,
% \begin{small}
% \begin{equation}
% \label{eq:MLP2}
% D(\mathbf{x}) = \sigma(\mathbf{W} \mathbf{x}+\mathbf{b}),
% \end{equation}
% \end{small}
where $D(\cdot)$ estimates probabilities of its inputs been real user profiles, $\mathbf{W}$ and $\mathbf{b}$ are weight matrix and bias vector.

Inspired by the idea of GAN~\citep{GoodfellowPMXWOCB14}, we aim to unify the different goals of generator and discriminator by letting them play a \emph{minimax game} via optimizing the following \emph{adversarial loss}: 
\begin{equation}
\label{eq:minmax}
\min_{\mathbf{\theta}}\max_{\mathbf{\phi}} H(G, D)=\mathbb{E}_{u\sim\mathcal{U}} [\log D(\mathbf{x}_u)]+\mathbb{E}_{z\sim p_{\mathbf{\theta}}}[\log\big(1-D(z)\big)],
\end{equation}
where $\mathbf{\theta}$ and $\mathbf{\phi}$ are model parameters of $G$ and $D$, respectively. $\mathbf{x}_u$ is a real user profile. $z$ is a fake user profile from the generator distribution $p_{\mathbf{\theta}}$. %, derived by \fix{$z=G(\mathbf{x}_{u'})$} where $u'\sim\mathcal{U'}$.

\subsection{Learning}
\label{sec:learning}
Finally, the complete objective considers adversarial loss, reconstruction loss and shilling loss, and leads to the following formulation: 
\begin{equation}
\label{equ:generator}
\mathcal{L}_{AUSH} = \min_{\mathbf{\theta}}\max_{\mathbf{\phi}} \Big( H(G, D)+ \mathcal{L}_{Shill} + \mathcal{L}_{Recon}\Big).
\end{equation}

As shown in Alg.~\ref{alg:training}, in each round of the optimization, each of the ``attacker'' (i.e., generator)
and ``defender'' (i.e., discriminator) endeavors to improve itself to defeat
the other part. The generator attempts to generate ``perfect'' fake user
profiles that are difficult to detect, while the discriminator tries to
accurately identify fake profiles. During this procedure, the generator learns
to produce fake profiles similar to real profiles via optimizing the
reconstruction loss. At the same time, optimizing the shilling loss endows
the fake profiles with the capability to exert ancillary influence (e.g.,
demote competitors or bias in-segment users).

\section{Experiment}
\label{sec:experiment}

In this section, we conduct experiments in order to answer the following research questions:
\begin{itemize}[leftmargin=14pt]
\item \textbf{RQ1}: Does AUSH have better attack performance on both traditional and deep learning based RS, than other shilling attack methods?
\item \textbf{RQ2}: If adversarial attack methods are directly used (i.e., using a general GAN) for shilling attack, what are the attack impacts?
\item \textbf{RQ3}: Is AUSH able to achieve secondary attack goals at the same time?
\item \textbf{RQ4}: How much does each component in AUSH contribute to the attack effects?
\item \textbf{RQ5}: Is it more difficult for attack detector to recognize the attack launched by AUSH, compared to shilling attack methods?
\end{itemize}

In the following, we first demonstrate our experiment setup in Sec.~\ref{sec:setup}. Then, the attack effect of AUSH is verified on three well-known recommendation benchmarks and is compared with both heuristic based and general GAN based attack models in Sec.~\ref{sec:attack} (RQ1, RQ2, RQ3). After that, we investigate the role of each component in AUSH on the attack impact (RQ4).
Finally, we show that AUSH can not be detected by supervised and unsupervised attack detection methods in Sec.~\ref{sec:detection} and it generates indistinguishable profiles in terms of similarity measurements (RQ5).

\subsection{Experimental Setup}\label{sec:setup}

We use three benchmark data sets for RS in our experiments: ML-100K\footnote{\url{https://grouplens.org/datasets/movielens/100k/}}, FilmTrust\footnote{\url{https://www.librec.net/datasets/filmtrust.zip}} and Amazon Automotive\footnote{\url{http://jmcauley.ucsd.edu/data/amazon/}}. Most of the previous work~\citep{SandvigMB08} only uses ML-100K as the single data set. We use its default training/test split. In addition, we use FilmTrust and Automotive, which are larger and sparser, to testify the competence of AUSH in different settings. We randomly split them by 9:1 for training and testing, respectively.
% Its ratings are mapped to $[1,5]$. Automotive contains the data of a subcategory in the collection of Amazon data.
To exclude cold-start users (as they are too vulnerable), we filter users with less than 15 ratings and items without ratings.

\begin{table}[t]
\caption{Statistics of data}
\begin{center}
\scalebox{0.75}{
\begin{tabular}{|>{\centering\arraybackslash}p{0.18\columnwidth}|>{\centering\arraybackslash}p{0.2\columnwidth}|>{\centering\arraybackslash}p{0.2\columnwidth}|>{\centering\arraybackslash}p{0.27\columnwidth}|>{\centering\arraybackslash}p{0.2\columnwidth}|}
\hline
Data      & \#Users     & \#Items     & \#Ratings        & Sparsity     \\ \hline
ML-100K   & 943         & 1,682       & 100,000          & 93.70\%       \\ \hline
FilmTrust & 780         & 721         & 28,799           & 94.88\%       \\ \hline
Automotive & 2,928		& 1,835		  & 20,473			 & 99.62\%			\\\hline
\end{tabular}
}
\newline
\vspace{5pt}
\newline
\scalebox{0.75}{
\begin{tabular}{|>{\centering\arraybackslash}p{0.18\columnwidth}|>{\centering\arraybackslash}p{0.2\columnwidth}|>{\centering\arraybackslash}p{0.2\columnwidth}|>{\centering\arraybackslash}p{0.27\columnwidth}|>{\centering\arraybackslash}p{0.2\columnwidth}|}
\hline
Data      & Attack Size & Filler Size & \#Selected Items & Profile Size \\ \hline
ML-100K   & 50          & 90          & 3                & 94           \\ \hline
FilmTrust & 50          & 35          & 2                & 38           \\ \hline
Automotive &50			& 4		      & 1		         & 6			\\\hline
\end{tabular}
}
\end{center}
\label{tab:datasets}
\end{table}

We inject $50$ user profiles (i.e., roughly 5\% of the population which can manifest the differences among attack models~\citep{BurkeMBW05}) in each attack. The number of
fillers in each injected user profile equals to the average number of ratings
per user in the data set.
%For target items, we select five \emph{random target items} from the item universe and five \emph {random long-tail target items} from items. Target items have no more than two ratings in the training set.
For each target item in ML-100K, we select a small number of items that are most frequently rated under the same tag/category of the target item as the selected items. For each target item in FilmTrust and Automotive which do not have information of tag/category, we sample items from global popular items as the selected items. Tab.~\ref{tab:datasets} illustrates the statistics of the data.

We use TensorFlow for the implementation. The generator of AUSH has 5 hidden layers (i.e., $N=5$) with $400$, $133$, $44$, $14$ and $4$ neurons for each layer. We use random sampling as the default strategy for sampling filler items in AUSH as it requires the least effort.
% By default, the complete loss is used in AUSH and $m$ is set to $50\%$ for reconstruction.
 % We report the results of other sampling strategies and impact of only using adversarial loss in Sec.~\ref{sec:parameters}.
The output layer size is the number of selected items. We use Adam~\cite{KingmaB14} for optimization with an initial learning rate of 0.01. The maximal number of adversarial iterations is set to be $150$.

\subsection{Attack Performance (RQ1, RQ2, RQ3)}\label{sec:attack}
To answer RQ1, we investigate the attack performance of AUSH and compare it with other baselines on several classic and deep learning based RS models including
NMF~\citep{Lee2001Algorithms}, NNMF~\citep{Dziugaite2015Neural} and AutoEncoder~\citep{Sedhain2015AutoRec} in our experiments.
Note that AUSH is designed for attacking rating based RS and we need to estimate the exact values of ratings in the experiment. Thus we exclude methods such as NCF~\citep{HeLZNHC17} which is designed for implicit feedback. We compare AUSH with several shilling attack models:
\begin{enumerate}[label=(\arabic*),leftmargin=15pt,topsep=3pt]
\item \textbf{Random attack} assigns a rating $r\sim\mathcal{N}(\mu,\sigma)$ to a filler, where $\mu$ and $\sigma$ are the mean and the variance of all ratings in the system, respectively.
\item \textbf{Average attack} assigns a rating $r\sim\mathcal{N}(\mu,\sigma)$ to a filler, where $\mu$ and $\sigma$ are the mean and the variance of ratings on this filler in the system, respectively.
\item \textbf{Segment attack} assigns maximal ratings to the selected items and minimal ratings to the filler items.
\item \textbf{Bandwagon attack} uses the most popular items as the selected items and assigns maximal ratings to them, while fillers are assigned ratings in the same manner as random attack.
\item \textbf{DCGAN} is an adversarial network~\citep{RadfordMC15} adopted in a recent shilling attack method~\citep{Christakopoulou2018Adversarial,Christakopoulou19}, where the generator takes the input noise and output fake user profiles through convolutional units. We use the default settings in~\citep{Christakopoulou19}.
\item \textbf{WGAN} is similar to DCGAN, but we replace the GAN used in the shilling attack with Wasserstein GAN~\citep{ArjovskyCB17} which has a good empirical performance~\citep{Ishaan2017WGAN}.
\end{enumerate}

In all methods, the highest rating is assigned to the target item.
We train each RS and AUSH until convergence.
The required information (e.g., mean
and variance) is obtained from the training set. Note the prior knowledge of AUSH does not exceed what the baselines require to know. Then we inject user profiles
generated by the attack models to the training set and train the
RS again on the polluted data. We evaluate the attack
performance on the test set using prediction
shift (PS) and Hit Ratios at $K$ (HR@$K$). PS is the difference of ratings that the RS makes before and after the attack. HR@$K$ is the hit
ratio of target items in the top-$K$ recommendations
after the
attack. As we are performing a push attack, the PS and HR@$K$
need to be positive to indicate an effective attack. The larger their values
are, the more effective the attack is. In the evaluation, we use $K=10$ for HR@$K$.

% and $m$ is set to $50\%$ for reconstruction
\vspace{5pt}
\noindent\textbf{Overall Performance (RQ1).} We randomly select five items as \emph{random targets}. As indicated in the literature~\citep{LamR04}, unpopular items (i.e., long-tail items) are likely to be the targets of an attack. Therefore, we additionally sample five target items with the number of ratings no more than a threshold as \emph{random long-tail targets}. The threshold number of ratings is one in ML-100K, two in FilmTrust, and three in Automotive. We report the average attack performance on the three data sets for random targets and random long-tail targets, when the complete loss $\mathcal{L}_{AUSH}$ (i.e., Eq.~\ref{equ:generator}) is used in AUSH, in Tabs.~\ref{tab:result1},~\ref{tab:result2},~\ref{tab:result3} and~\ref{tab:result4} for attacking NMF, NNMF, U-AutoEncoder and I-AutoEncoder, respectively. We highlight the best performance in each category.
AUSH will also be highlighted if it achieves the second best performance.
% All
% results are averaged over target items.

From experimental results, we can see
that \emph{AUSH generally achieves attractive attack performance against all
recommendation models} on target items. It generates the largest PS and HR@$10$
in most categories including both random targets and random long-tail targets, showing that AUSH is a practical
method for attackers who want to promote their products. %
\emph{Conventional attack models do not show a robust attack
performance like AUSH, even though they may exceed AUSH in a few cases.}

\vspace{5pt}
\noindent\textbf{Comparisons between AUSH and General GANs (RQ2).}
We can observe from Tabs.~\ref{tab:result1},~\ref{tab:result2},~\ref{tab:result3} and~\ref{tab:result4} that \emph{directly adopting the idea of adversarial attacks (i.e., using general GANs) does not give a satisfactory performance}. Particularly, both DCGAN which is adopted in the recent shilling attack~\citep{Christakopoulou2018Adversarial,Christakopoulou19} and WGAN~\citep{ArjovskyCB17} which aims at stabilizing GAN training do not show better performance than simple heuristic based attack approaches like Average attack and Random attack. In some cases, attacks launched by DCGAN and WGAN even give opposite effects (i.e., negative PS and HR@$K$). It validates our assumption that a tailored GAN framework is necessary for shilling attack.

\vspace{5pt}
\noindent\textbf{Secondary Attack Goals (RQ3).}
As explained in Sec.~\ref{sec:generator}, incorporating a shilling loss helps AUSH to achieve secondary attack goal, i.e., increasing the impact of the attack on users who like the selected items before the attack.
% In-segment users are users who have shown preferences on selected items in the training set.
%We define in-segment users as users who have assigned high
%ratings (i.e., $4$- or $5$-stars) on all selected item in the
%training set.
We call such users \emph{in-segment users}~\citep{LamR04} and they are target population to certain attackers. We define
in-segment users as users who have assigned high ratings
(i.e., 4- or 5-stars) on all selected item in our experiments.
In Tabs.~\ref{tab:result1},~\ref{tab:result2},~\ref{tab:result3} and~\ref{tab:result4}, we already report the attack results for in-segment users and all users, and list random targets and random long-tail targets separately.
In Tab.~\ref{tab:result_sample}, we further show the attack performance on I-AutoEncoder in different settings and the results are reported by averaging attack impacts on random targets and random long-tail targets. Note that AUSH$_{rand}$ in Tab.~\ref{tab:result_sample} indicates the complete loss (i.e., Eq.~\ref{equ:generator}) and the default random sampling are used, i.e., it is equivalent to AUSH in Tabs.~\ref{tab:result1},~\ref{tab:result2},~\ref{tab:result3} and~\ref{tab:result4}. For example, AUSH$_{rand}$ has a PS of 1.6623 for in-segment users in Tab.~\ref{tab:result_sample} which is the average of AUSH's attack performances for in-segment users on random targets and random long-tail targets of ML-100K (i.e., 1.3746 and 1.9499) in Tab.~\ref{tab:result4}.
Due to space limit, we only report attack results in ML-100K in Tab.~\ref{tab:result_sample}, but we observe similar results in other settings. 

We can observe that AUSH (in Tabs.~\ref{tab:result1},~\ref{tab:result2},~\ref{tab:result3} and~\ref{tab:result4}) and AUSH$_{rand}$ (in Tab.~\ref{tab:result_sample}) enhance the power of segment attack -- a much more significant attack impact on in-segment users than on all users, while other baselines are not that flexible and they are unable to achieve such a secondary attack goal. This property of AUSH is desirable if the attacker wants to demote the items from competitors. Note AUSH$_{rand}$ uses the complete loss.
For a further study on impacts of the different loss components, please refer to the next section.

% \fix{\emph{AUSH can simultaneously achieve multiple attack goals}.
% We distinguish in-segment users from all users, long-tail target items from random target items.
% We can see that AUSH is effective on in-segment users, which is desirable if the attacker wants to strke competitors.
% AUSH is also a practical method for attackers who want to promote their new products with very little historical information.}

\begin{table*}[!t]
\caption{Attack performance against NMF. Best results are marked in bold, and AUSH results are also marked in bold if they are the second best in each category.}
\label{tab:result1}
\resizebox{0.95\textwidth}{!}{
\begin{tabular}{|c|l|c|c|c|c|c|c|c|c|c|c|c|c|}
\cline{1-1} \cline{3-14}
\multirow{2}{*}{Metric} &                       & \multicolumn{6}{c|}{In-segment Users}                                                                     & \multicolumn{6}{c|}{All Users}                                                                            \\ \cline{3-14}
                        &                       & \multicolumn{3}{c|}{Prediction Shift}               & \multicolumn{3}{c|}{HR@10}                          & \multicolumn{3}{c|}{Prediction Shift}               & \multicolumn{3}{c|}{HR@10}                          \\ \cline{1-1} \cline{3-14}
Data Set                 &                       & ML-100K         & FilmTrust       & Automotive      & ML-100K         & FilmTrust       & Automotive      & ML-100K         & FilmTrust       & Automotive      & ML-100K         & FilmTrust       & Automotive      \\ \cline{1-1} \cline{3-14}
Model                   &                       & \multicolumn{12}{c|}{Random Targets}                                                                                                                                                                                  \\ \cline{1-1} \cline{3-14}
AUSH                    &                       & \textbf{1.8857} & \textbf{0.8937} & \textbf{0.2778} & \textbf{0.2538} & \textbf{0.2822} & \textbf{0.0539} & \textbf{1.7503} & \textbf{0.9650} & \textbf{0.2585} & \textbf{0.1849} & \textbf{0.2821} & \textbf{0.0541}
 \\ \cline{1-1} \cline{3-14}
Segment                 &                       & 1.0157          & 0.6832          & 0.2313          & 0.0372          & 0.3214          & \textbf{0.1545} & 0.7061          & 0.4504          & \textbf{0.2649} & 0.0380          & 0.1978          & \textbf{0.1132} \\ \cline{1-1} \cline{3-14}
Average                 &                       &1.8478          & 0.8721          & 0.1972          & 0.2147          & 0.2208          & 0.0239          & 1.7754          & 0.9522          & 0.2100          & 0.1787          & 0.2241          & 0.0241          \\ \cline{1-1} \cline{3-14}
Random                  &                       & 1.7220          & 0.8667          & 0.2332          & 0.1253          & 0.2708          & 0.0380          & 1.6285          & 0.9570          & 0.2391          & 0.0995          & 0.3140          & 0.0406          \\ \cline{1-1} \cline{3-14}
Bandwagon               &                       & 1.7199          & 0.8184          & 0.2380          & 0.1791          & 0.2380          & 0.0294          & 1.6194          & 0.8508          & 0.2327          & 0.1257          & 0.2048          & 0.0300          \\ \cline{1-1} \cline{3-14}
DCGAN                   & \multicolumn{1}{c|}{} & -0.0112         & 0.1082          & 0.1002          & 0.0000          & 0.0833          & 0.0086          & -0.0096         & 0.1005          & 0.1065          & 0.0000          & 0.0751          & 0.0046          \\ \cline{1-1} \cline{3-14}
WGAN                    & \multicolumn{1}{c|}{} & 0.0774          & 0.1966          & 0.0473          & 0.0000          & 0.0469          & 0.0040          & 0.0723          & 0.1923          & 0.0396          & 0.0000          & 0.0374          & 0.0055          \\ \cline{1-1} \cline{3-14}
                        &                       & \multicolumn{12}{c|}{Random Long-tail Targets}                                                                                                                                                                        \\ \cline{1-1} \cline{3-14}
AUSH                    &                       & \textbf{2.9387} & \textbf{1.4263} & \textbf{0.2575} & \textbf{0.6007} & \textbf{0.1571} & 0.0055          & \textbf{2.8949} & \textbf{1.4758} & 0.2456          & \textbf{0.5057} & \textbf{0.1961} & \textbf{0.0091} \\ \cline{1-1} \cline{3-14}

Segment                 &                       & 2.7918          & 0.9993          & 0.1719          & 0.5175          & 0.2197          & \textbf{0.0669} & 2.5726          & 0.7095          & \textbf{0.2961} & 0.3450          & 0.1265          & \textbf{0.0541} \\ \cline{1-1} \cline{3-14}
Average                 &                       & 2.9427          & 1.4084          & 0.2508          & 0.5044          & 0.0941          & 0.0066          & 2.9038          & 1.4723          & 0.2544          & 0.4420          & 0.1247          & 0.0041          \\ \cline{1-1} \cline{3-14}
Random                  &                       & 2.8994          & 1.4084          & \textbf{0.2618} & 0.6661          & 0.1568          & 0.0050          & 2.8401          & 1.4718          & 0.2724          & 0.5276          & 0.2159          & 0.0091          \\ \cline{1-1} \cline{3-14}
Bandwagon               &                       & 2.8752          & 1.3426          & 0.1385          & 0.6232          & 0.1412          & 0.0000          & 2.8100          & 1.3561          & 0.1628          & 0.4900          & 0.1501          & 0.0011          \\ \cline{1-1} \cline{3-14}
DCGAN                   & \multicolumn{1}{c|}{} & -0.1479         & 0.1753          & -0.0731         & 0.0000          & 0.0008          & 0.0000          & -0.1374         & 0.1836          & -0.0383         & 0.0000          & 0.0088          & 0.0002          \\ \cline{1-1} \cline{3-14}
WGAN                    & \multicolumn{1}{c|}{} & 1.2299          & 0.4455          & -0.0509         & 0.0000          & 0.0332          & 0.0000          & 1.2473          & 0.4071          & -0.0416         & 0.0000          & 0.0298          & 0.0016          \\ \cline{1-1} \cline{3-14}
\end{tabular}}
\end{table*}

% \usepackage{multirow}
\begin{table*}[!t]
\caption{Attack performance against NNMF. Best results are marked in bold, and AUSH results are also marked in bold if they are the second best in each category.}
\label{tab:result2}
\resizebox{0.95\textwidth}{!}{
\begin{tabular}{|c|l|c|c|c|c|c|c|c|c|c|c|c|c|}
\cline{1-1} \cline{3-14}
\multirow{2}{*}{Metric} &                       & \multicolumn{6}{c|}{In-segment Users}                                                                      & \multicolumn{6}{c|}{All Users}                                                                             \\ \cline{3-14}
                        &                       & \multicolumn{3}{c|}{Prediction Shift}               & \multicolumn{3}{c|}{HR@10}                           & \multicolumn{3}{c|}{Prediction Shift}               & \multicolumn{3}{c|}{HR@10}                           \\ \cline{1-1} \cline{3-14}
Data Set                 &                       & ML-100K         & FilmTrust       & Automotive      & ML-100K         & FilmTrust       & Automotive       & ML-100K         & FilmTrust       & Automotive      & ML-100K         & FilmTrust       & Automotive       \\ \cline{1-1} \cline{3-14}
Model                   &                       & \multicolumn{12}{c|}{Random Targets}                                                                                                                                                                                    \\ \cline{1-1} \cline{3-14}
AUSH                    &                       & \textbf{1.2225} & \textbf{0.9092} & 0.2507          & \textbf{0.1170} & \textbf{0.3027} & 0.0242          & \textbf{1.4009} & \textbf{1.1156} & 0.3017          & \textbf{0.1704} & \textbf{0.3614} & 0.0254          \\ \cline{1-1} \cline{3-14}
Segment                 &                       & 0.0500          & 0.4423          & 0.1745          & 0.0156          & 0.1330          & 0.0213          & -0.4469         & 0.4486          & 0.1701          & 0.0069          & 0.1240          & 0.0242          \\ \cline{1-1} \cline{3-14}
Average                 &                       & 0.8749          & 0.7795          & \textbf{0.3016} & 0.0665          & 0.2220          & 0.0279          & 1.1468          & 0.9129          & \textbf{0.3491} & 0.1112          & 0.2340          & 0.0392          \\ \cline{1-1} \cline{3-14}
Random                  &                       & 0.5837          & 0.7634          & 0.2815          & 0.0431          & 0.1568          & \textbf{0.0399} & 0.8732          & 0.9334          & 0.3005          & 0.0411          & 0.2083          & \textbf{0.0426} \\ \cline{1-1} \cline{3-14}
Bandwagon               &                       & 0.6517          & 0.7333          & 0.2716          & 0.0388          & 0.1945          & 0.0223          & 0.5153          & 0.8634          & 0.3157          & 0.0309          & 0.2168          & 0.0260          \\ \cline{1-1} \cline{3-14}
DCGAN                   & \multicolumn{1}{c|}{} & -0.0611         & -0.2444         & 0.0468          & 0.0012          & 0.0010          & 0.0000           & 0.0885          & -0.1889         & 0.0274          & 0.0013          & 0.0034          & 0.0010           \\ \cline{1-1} \cline{3-14}
WGAN                    & \multicolumn{1}{c|}{} & -0.0543         & 0.0786          & 0.0093          & 0.0000          & 0.0600          & 0.0100           & -0.0649         & 0.1085          & -0.0041         & 0.0007          & 0.0457          & 0.0037           \\ \cline{1-1} \cline{3-14}
                        &                       & \multicolumn{12}{c|}{Random Long-tail Targets}                                                                                                                                                                          \\ \cline{1-1} \cline{3-14}
AUSH                    &                       & \textbf{1.5956} & \textbf{0.9002} & \textbf{0.8406} & \textbf{0.2654} & \textbf{0.2957} & \textbf{0.0257} & \textbf{1.7413} & \textbf{1.1241} & \textbf{0.8343} & \textbf{0.3420} & \textbf{0.3799} & \textbf{0.0206} \\ \cline{1-1} \cline{3-14}
Segment                 &                       & -0.4232         & 0.3003          & 0.5454          & 0.0011          & 0.1360          & 0.0116          & -0.8599         & 0.3996          & 0.5150          & 0.0011          & 0.1242          & 0.0162          \\ \cline{1-1} \cline{3-14}
Average                 &                       & 1.4323          & 0.7883          & 0.8203          & 0.1503          & 0.1532          & 0.0188          & 1.5251          & 0.9430          & 0.7721          & 0.2236          & 0.1841          & 0.0158          \\ \cline{1-1} \cline{3-14}
Random                  &                       & 1.3755          & 0.8430          & 0.8023          & 0.1432          & 0.2011          & \textbf{0.0307} & 1.4984          & 1.0222          & 0.7878          & 0.2255          & 0.2648          & \textbf{0.0402} \\ \cline{1-1} \cline{3-14}
Bandwagon               &                       & 1.3315          & 0.6977          & 0.5278          & 0.1296          & 0.1143          & 0.0102          & 1.4923          & 0.8026          & 0.5131          & 0.1877          & 0.1306          & 0.0056          \\ \cline{1-1} \cline{3-14}
DCGAN                   & \multicolumn{1}{c|}{} & 0.1487          & -0.4251         & 0.1673          & 0.0000          & 0.0010          & 0.0000           & 0.2164          & -0.3518         & 0.1438          & 0.0000          & 0.0008          & 0.0028           \\ \cline{1-1} \cline{3-14}
WGAN                    & \multicolumn{1}{c|}{} & 0.0555          & 0.1383          & 0.2385          & 0.0000          & 0.0021          & 0.0000           & 0.0266          & 0.2591          & 0.1144          & 0.0000          & 0.0033          & 0.0081           \\ \cline{1-1} \cline{3-14}
\end{tabular}}
\end{table*}

\subsection{Impacts of Sampling Strategies and Each Loss (RQ4)}\label{sec:lossimpact}
To answer RQ4, we remove or change some components of AUSH and investigate the performance changes.

\vspace{5pt}
\noindent\textbf{Impacts of Sampling Strategies.}
We report the impacts of different sampling strategies in
Tab.~\ref{tab:result_sample}. AUSH$_{rand}$,
AUSH$_{rating}$, AUSH$_{pop}$ and AUSH$_{sim}$ indicate random sample, sample by
rating, sample by popularity and sample by similarity, respectively. All the four variations of AUSH adopt the complete loss (i.e., Eq.~\ref{equ:generator}).
We can observe that sample by rating is the
best strategy for AUSH. The reason may be that it is easy to bias
people with items having high ratings, as customers tend to trust such
``power'' items~\citep{SeminarioW14b}. Nevertheless, all the variations have more significant attack impacts than other baselines.

\vspace{5pt}
\noindent\textbf{Impacts of Each Loss.}
To study the contributions of each loss term, in
Tab.~\ref{tab:result_sample}, we also report the results of AUSH$_{adv}$, AUSH$_{rec}$ and AUSH$_{rec+shill}$, which denote using adversarial loss only, using reconstruction loss only, and using reconstruction and shilling losses, respectively. In these three methods, random sampling is employed.
An ordinary neural network (i.e., AUSH$_{rec}$) is outperformed by the complete AUSH (i.e., AUSH$_{rand}$, AUSH$_{rating}$, AUSH$_{pop}$ and AUSH$_{sim}$), showing the effectiveness of our design of tailoring GAN for use in shilling attacks.
AUSH$_{adv}$ has the
worst attack performance compared to other variations of AUSH, showing that the reconstruction loss also contributes to the attack.

\begin{table*}[!t]
\caption{Attack performance against U-AutoEncoder. Best results are marked in bold, and AUSH results are also marked in bold if they are the second best in each category.}
\label{tab:result3}
\resizebox{0.95\textwidth}{!}{
\begin{tabular}{|c|l|c|c|c|c|c|c|c|c|c|c|c|c|}
\cline{1-1} \cline{3-14}
                         &                       & \multicolumn{6}{c|}{In-segment Users}                                                                                                                                                                              & \multicolumn{6}{c|}{All Users}                                                                                                                                                                                     \\ \cline{3-14}
\multirow{-2}{*}{Metric} &                       & \multicolumn{3}{c|}{Prediction Shift}                                                                   & \multicolumn{3}{c|}{HR@10}                                                                               & \multicolumn{3}{c|}{Prediction Shift}                                                                     & \multicolumn{3}{c|}{HR@10}                                                                             \\ \cline{1-1} \cline{3-14}
Data Set                  &                       & ML-100K                        & FilmTrust                     & \multicolumn{1}{l|}{Automotive}        & \multicolumn{1}{l|}{ML-100K}  & \multicolumn{1}{l|}{FilmTrust}         & \multicolumn{1}{l|}{Automotive} & \multicolumn{1}{l|}{ML-100K}   & \multicolumn{1}{l|}{FilmTrust} & \multicolumn{1}{l|}{Automotive}         & \multicolumn{1}{l|}{ML-100K} & \multicolumn{1}{l|}{FilmTrust} & \multicolumn{1}{l|}{Automotive}        \\ \cline{1-1} \cline{3-14}
Model                    &                       & \multicolumn{12}{c|}{Random Targets}                                                                                                                                                                                                                                                                                                                                                                                                    \\ \cline{1-1} \cline{3-14}
AUSH                     &                       & \textbf{1.7661}                & \textbf{1.3406}               & \textbf{0.2206}                        & \textbf{0.2465}               & \textbf{0.5596}                        & \textbf{0.0168}                 & \textbf{1.6184}                & \textbf{1.1550}                & \textbf{0.0382}                         & \textbf{0.2006}              & \textbf{0.3549}                & \textbf{0.0050}                        \\ \cline{1-1} \cline{3-14}
Segment                  &                       & 0.4721                         & 1.0875                        & \textbf{0.4700}                        & 0.0036                        & 0.5371                                 & \textbf{0.7789}                 & 0.3098                         & 0.8886                         & 0.0121                                  & 0.0050                       & \textbf{0.3719}                & \textbf{0.2166}                        \\ \cline{1-1} \cline{3-14}
Average                  &                       & 0.9297                         & 0.9024                        & 0.1311                                 & 0.0144                        & 0.1490                                 & 0.0000                          & 1.0187                         & 0.9731                         & 0.1514                                  & 0.0231                       & 0.1481                         & 0.0000                                 \\ \cline{1-1} \cline{3-14}
Random                   &                       & 0.4624                         & 0.7527                        & 0.1262                                 & 0.0027                        & 0.0807                                 & 0.0000                          & 0.6284                         & 0.8271                         & 0.1200                                  & 0.0059                       & 0.1023                         & 0.0000                                 \\ \cline{1-1} \cline{3-14}
Bandwagon                &                       & 0.5501                         & 0.6026                        & 0.0896                                 & 0.0012                        & 0.0316                                 & 0.0000                          & 0.6311                         & 0.6382                         & 0.0686                                  & 0.0062                       & 0.0335                         & 0.0000                                 \\ \cline{1-1} \cline{3-14}
DCGAN                    & \multicolumn{1}{c|}{} & -1.064                         & {\color[HTML]{000000} 0.0076} & {\color[HTML]{000000} -0.2258}         & {\color[HTML]{000000} 0.0000} & {\color[HTML]{000000} 0.0000}          & {\color[HTML]{000000} 0.0000}   & {\color[HTML]{000000} -0.2215} & {\color[HTML]{000000} -0.0326} & -0.2415                                 & 0.0000                       & 0.0000                         & 0.0000                                 \\ \cline{1-1} \cline{3-14}
WGAN                     & \multicolumn{1}{c|}{} & {\color[HTML]{000000} 1.3940}  & {\color[HTML]{000000} 0.0923} & {\color[HTML]{000000} 0.1813}          & {\color[HTML]{000000} 0.0000} & {\color[HTML]{000000} 0.0000}          & {\color[HTML]{000000} 0.1583}   & {\color[HTML]{000000} 1.2985}  & {\color[HTML]{000000} 0.1095}  & {\color[HTML]{000000} \textbf{0.1630}}  & 0.0212                       & {\color[HTML]{000000} 0.0000}  & 0.0928                                 \\ \cline{1-1} \cline{3-14}
                         &                       & \multicolumn{12}{c|}{Random Long-tail Targets}                                                                                                                                                                                                                                                                                                                                                                                          \\ \cline{1-1} \cline{3-14}
AUSH                     &                       & \textbf{3.2274}                & \textbf{1.7384}               & {\color[HTML]{333333} \textbf{0.3898}} & \textbf{0.6657}               & {\color[HTML]{000000} \textbf{0.6896}} & \textbf{0.0000}                 & \textbf{2.9440}                & \textbf{1.5602}                & {\color[HTML]{000000} \textbf{-0.0424}} & \textbf{0.4894}              & \textbf{0.5149}                & \textbf{0.0000}                        \\ \cline{1-1} \cline{3-14}
Segment                  &                       & \textbf{3.3397}                & 1.4665                        & {\color[HTML]{000000} 0.2109}          & 0.6364                        & 0.5570                                 & \textbf{0.3733}                 & 3.0081                         & 1.2709                         & -0.4654                                 & \textbf{0.5423}              & 0.4175                         & 0.0098                                 \\ \cline{1-1} \cline{3-14}
Average                  &                       & 3.1671                         & 1.2961                        & {\color[HTML]{000000} 0.2915}          & 0.3897                        & 0.0425                                 & 0.0000                          & \textbf{3.0299}                & 1.3290                         & \textbf{0.2930}                         & 0.4439                       & 0.0851                         & 0.0000                                 \\ \cline{1-1} \cline{3-14}
Random                   &                       & 2.5778                         & 1.0348                        & {\color[HTML]{000000} 0.0466}          & 0.1508                        & 0.0259                                 & 0.0000                          & 2.5229                         & 1.1324                         & 0.0275                                  & 0.1575                       & 0.0815                         & 0.0000                                 \\ \cline{1-1} \cline{3-14}
Bandwagon                &                       & 2.5466                         & 0.8524                        & {\color[HTML]{000000} 0.1227}          & 0.1581                        & 0.0073                                 & 0.0000                          & 2.4444                         & 0.9117                         & 0.0509                                  & 0.1242                       & 0.0198                         & 0.0000                                 \\ \cline{1-1} \cline{3-14}
DCGAN                    & \multicolumn{1}{c|}{} & {\color[HTML]{000000} -0.3896} & {\color[HTML]{000000} 0.3782} & {\color[HTML]{000000} -0.0539}         & {\color[HTML]{000000} 0.0000} & {\color[HTML]{000000} 0.0000}          & {\color[HTML]{000000} 0.0000}   & {\color[HTML]{000000} -0.3813} & {\color[HTML]{000000} 0.4132}  & 0.0496                                  & 0.0000                       & 0.0000                         & 0.0000                                 \\ \cline{1-1} \cline{3-14}
WGAN                     & \multicolumn{1}{c|}{} & {\color[HTML]{000000} 1.3940}  & {\color[HTML]{000000} 0.0923} & {\color[HTML]{000000} 0.1813}          & {\color[HTML]{000000} 0.0000} & {\color[HTML]{000000} 0.0000}          & {\color[HTML]{000000} 0.1583}   & {\color[HTML]{000000} 1.2985}  & {\color[HTML]{000000} 0.1095}  & {\color[HTML]{000000} 0.1630}           & 0.0212                       & 0.0000                         & {\color[HTML]{000000} \textbf{0.0928}} \\ \cline{1-1} \cline{3-14}
\end{tabular}}
\end{table*}

\begin{table*}[!t]
\caption{Attack performance against I-AutoEncoder. Best results are marked in bold, and AUSH results are also marked in bold if they are the second best in each category.}
\label{tab:result4}
\resizebox{0.95\textwidth}{!}{
\begin{tabular}{|c|l|c|c|c|c|c|c|c|c|c|c|c|c|}
\cline{1-1} \cline{3-14}
                         &                       & \multicolumn{6}{c|}{In-segment Users}                                                                                                                                                               & \multicolumn{6}{c|}{All Users}                                                                                                                                                                     \\ \cline{3-14}
\multirow{-2}{*}{Metric} &                       & \multicolumn{3}{c|}{Prediction Shift}                                                            & \multicolumn{3}{c|}{HR@10}                                                                       & \multicolumn{3}{c|}{Prediction Shift}                                                            & \multicolumn{3}{c|}{HR@10}                                                                      \\ \cline{1-1} \cline{3-14}
Data Set                  &                       & ML-100K                       & FilmTrust                      & \multicolumn{1}{l|}{Automotive} & \multicolumn{1}{l|}{ML-100K}  & \multicolumn{1}{l|}{FilmTrust} & \multicolumn{1}{l|}{Automotive} & \multicolumn{1}{l|}{ML-100K}  & \multicolumn{1}{l|}{FilmTrust} & \multicolumn{1}{l|}{Automotive} & \multicolumn{1}{l|}{ML-100K} & \multicolumn{1}{l|}{FilmTrust} & \multicolumn{1}{l|}{Automotive} \\ \cline{1-1} \cline{3-14}
Model                    &                       & \multicolumn{12}{c|}{Random Targets}                                                                                                                                                                                                                                                                                                                                                                     \\ \cline{1-1} \cline{3-14}
AUSH                     &                       & \textbf{1.3746}               & \textbf{1.4280}                & \textbf{0.9913}                 & \textbf{0.1488}               & \textbf{0.9155}                & \textbf{0.9141}                 & \textbf{1.2180}               & \textbf{1.3059}                & \textbf{0.8870}                 & \textbf{0.0990}              & \textbf{0.8333}                & \textbf{0.8965}                 \\ \cline{1-1} \cline{3-14}
Segment                  &                       & 0.5137                        & 1.2035                         & 0.4927                          & 0.0086                        & 0.6423                         & 0.7266                          & 0.3232                        & 0.8689                         & 1.6986                          & 0.0274                       & 0.4371                         & \textbf{0.9777}                 \\ \cline{1-1} \cline{3-14}
Average                  &                       & 1.0117                        & 1.4203                         & 0.5394                          & 0.0487                        & \textbf{0.9187}                & 0.6490                          & 1.1044                        & \textbf{1.3589}                & 0.5470                          & \textbf{0.1025}              & \textbf{0.8520}                & 0.6500                          \\ \cline{1-1} \cline{3-14}
Random                   &                       & 0.6304                        & 1.2210                         & 0.5492                          & 0.0585                        & 0.8307                         & 0.6732                          & 0.7634                        & 1.1630                         & 0.5391                          & 0.0918                       & 0.7477                         & 0.6483                          \\ \cline{1-1} \cline{3-14}
Bandwagon                &                       & 0.5978                        & 1.2788                         & 0.9718                          & 0.0287                        & 0.8299                         & 0.8608                          & 0.5960                        & 1.2297                         & \textbf{1.8099}                 & 0.0430                       & 0.7825                         & 0.9309                          \\ \cline{1-1} \cline{3-14}
DCGAN                    & \multicolumn{1}{c|}{} & 0.0243                        & {\color[HTML]{000000} -0.0633} & {\color[HTML]{000000} 0.0046}   & {\color[HTML]{000000} 0.0000} & {\color[HTML]{000000} 0.0010}  & {\color[HTML]{000000} 0.0050}   & {\color[HTML]{000000} 0.0213} & {\color[HTML]{000000} -0.0600} & 0.0054                          & 0.0000                       & 0.0010                         & 0.0054                          \\ \cline{1-1} \cline{3-14}
WGAN                     & \multicolumn{1}{c|}{} & {\color[HTML]{000000} 0.1131} & {\color[HTML]{000000} -0.1228} & {\color[HTML]{000000} 0.0412}   & {\color[HTML]{000000} 0.0000} & {\color[HTML]{000000} 0.0000}  & {\color[HTML]{000000} 0.0050}   & {\color[HTML]{000000} 0.1045} & {\color[HTML]{000000} -0.1142} & 0.0465                          & 0.0002                       & 0.0008                         & 0.0047                          \\ \cline{1-1} \cline{3-14}
                         &                       & \multicolumn{12}{c|}{Random Long-tail Targets}                                                                                                                                                                                                                                                                                                                                                           \\ \cline{1-1} \cline{3-14}
AUSH                     &                       & \textbf{1.9499}               & \textbf{1.7052}                & \textbf{0.9820}                 & \textbf{0.2974}               & \textbf{0.9239}                & \textbf{0.8821}                 & \textbf{1.7822}               & \textbf{1.6019}                & \textbf{0.8150}                 & 0.2369                       & 0.8396                         & \textbf{0.8623}                 \\ \cline{1-1} \cline{3-14}
Segment                  &                       & 0.5188                        & 1.4510                         & 0.3969                          & 0.0072                        & 0.5835                         & 0.5938                          & 0.3249                        & 1.1385                         & \textbf{1.5154}                 & 0.0344                       & 0.4165                         & \textbf{0.9629}                 \\ \cline{1-1} \cline{3-14}
Average                  &                       & 1.3898                        & 1.6790                         & 0.4245                          & 0.1019                        & 0.9041                         & 0.3726                          & 1.3793                        & \textbf{1.6318}                & 0.4478                          & 0.1104                       & \textbf{0.8483}                & 0.3846                          \\ \cline{1-1} \cline{3-14}
Random                   &                       & 0.9227                        & 1.4590                         & 0.46697                         & 0.0401                        & 0.7768                         & 0.4368                          & 1.0349                        & 1.4076                         & 0.4740                          & 0.0900                       & 0.6910                         & 0.4477                          \\ \cline{1-1} \cline{3-14}
Bandwagon                &                       & 0.6220                        & 1.5672                         & 0.2814                          & 0.0091                        & 0.8390                         & 0.4267                          & 0.7456                        & 1.5190                         & 0.6489                          & 0.0346                       & 0.7728                         & 0.6593                          \\ \cline{1-1} \cline{3-14}
DCGAN                    & \multicolumn{1}{c|}{} & {\color[HTML]{000000} 0.0241} & {\color[HTML]{000000} 0.0119}  & {\color[HTML]{000000} 0.0056}   & {\color[HTML]{000000} 0.0000} & {\color[HTML]{000000} 0.0000}  & {\color[HTML]{000000} 0.0000}   & {\color[HTML]{000000} 0.0348} & {\color[HTML]{000000} 0.0114}  & -0.0029                         & 0.000                        & 0.0005                         & 0.0086                          \\ \cline{1-1} \cline{3-14}
WGAN                     & \multicolumn{1}{c|}{} & {\color[HTML]{000000} 0.1096} & {\color[HTML]{000000} 0.0718}  & {\color[HTML]{000000} -0.0428}  & {\color[HTML]{000000} 0.0000} & {\color[HTML]{000000} 0.0000}  & {\color[HTML]{000000} 0.0000}   & {\color[HTML]{000000} 0.1374} & {\color[HTML]{000000} 0.0728}  & -0.0364                         & 0.0006                       & 0.0003                         & 0.0018                          \\ \cline{1-1} \cline{3-14}
\end{tabular}}
\end{table*}

\begin{table}[!t]
\caption{Attack performance on I-AutoEncoder using different sampling strategies and losses in ML-100K. Best results are marked in bold.}
\label{tab:result_sample}
\begin{center}
\resizebox{0.96\columnwidth}{!}{
\begin{tabular}{|l|c|c|c|c|}
\hline
Attack  Method  & \multicolumn{2}{c|}{In-segment Users}                              & \multicolumn{2}{c|}{All Users}                                     \\ \hline
                & \multicolumn{1}{l|}{Prediction Shift} & \multicolumn{1}{l|}{HR@10} & \multicolumn{1}{l|}{Prediction Shift} & \multicolumn{1}{l|}{HR@10} \\ \hline
AUSH$_{rand}$   &1.6623 & 0.2231 & 1.5001  & 0.1679                    \\ \hline
AUSH$_{rating}$ & \textbf{1.7310}                       & \textbf{0.2735}            & \textbf{1.5695}                       & \textbf{0.2243}            \\ \hline
AUSH$_{pop}$    & 1.7252                                & 0.2699                     & 1.5620                                & 0.2212                     \\ \hline
AUSH$_{sim}$    & 1.6752                                & 0.2300                     & 1.5383                                & 0.1992                     \\ \hline
AUSH$_{adv}$  & 1.2960                                & 0.0640                     & 1.3162                                & 0.0881                     \\ \hline
AUSH$_{rec}$                & 1.4980                                & 0.1450                     & 1.4411                                & 0.1441                     \\ \hline
AUSH$_{rec+shill}$          & 1.6569                                & 0.2349                     & 1.5033                                & 0.1849                     \\ \hline
Segment         & 0.5163 & 0.0079 & 0.3241 & 0.0309                     \\ \hline
Average         & 1.2008 & 0.0753 & 1.2419 & 0.1065                     \\ \hline
Random          & 0.7766 & 0.0493 & 0.8992 & 0.0909                     \\ \hline
Bandwagon       & 0.6099  & 0.0189 & 0.6708  & 0.0388                     \\ \hline
DCGAN       & 0.0242                               & 0.0000                     & 0.0281                                & 0.0000                     \\ \hline
WGAN       & 0.1114                               & 0.0000                     & 0.1210                                & 0.0004                     \\ \hline
\end{tabular}
}
\end{center}
\end{table}

\begin{table}[!t]
\caption{Two distance measures between injected user profiles and real user profiles in ML-100K.}
\label{tab:similarity}
\begin{center}
\resizebox{0.98\columnwidth}{!}{
% \begin{tabular}{|c|c|c|c|c|c|c|}
% \hline
% method	&AUSH	& Aush$_{rec}$ &	average	& bandwagon &	random	& segment \\\hline
% TVD	& \textbf{0.01210}	& 0.01213 	&	0.05450		&0.05762		&0.05704	&	0.08010\\\hline
% JS &	\textbf{0.00215}&	0.00219	&	0.01162		&0.01398		&0.01353		&0.03461\\\hline
% \end{tabular}
\begin{tabular}{|c|c|c|c|c|c|c|c|}
\hline
Measure	&AUSH	&	Average	& Bandwagon &	Random	& Segment & DCGAN & WGAN\\\hline
TVD	& \textbf{0.01210}		&	0.05450		&0.05762		&0.05704	&0.08010	&0.11302	&0.11598\\\hline
JS &	\textbf{0.00215}	&	0.01162		&0.01398		&0.01353    &0.03461	&0.04363	&0.04601\\\hline
\end{tabular}
}
\end{center}
\end{table}

\begin{figure}[t]
\begin{center}
\includegraphics[width=0.99\columnwidth]{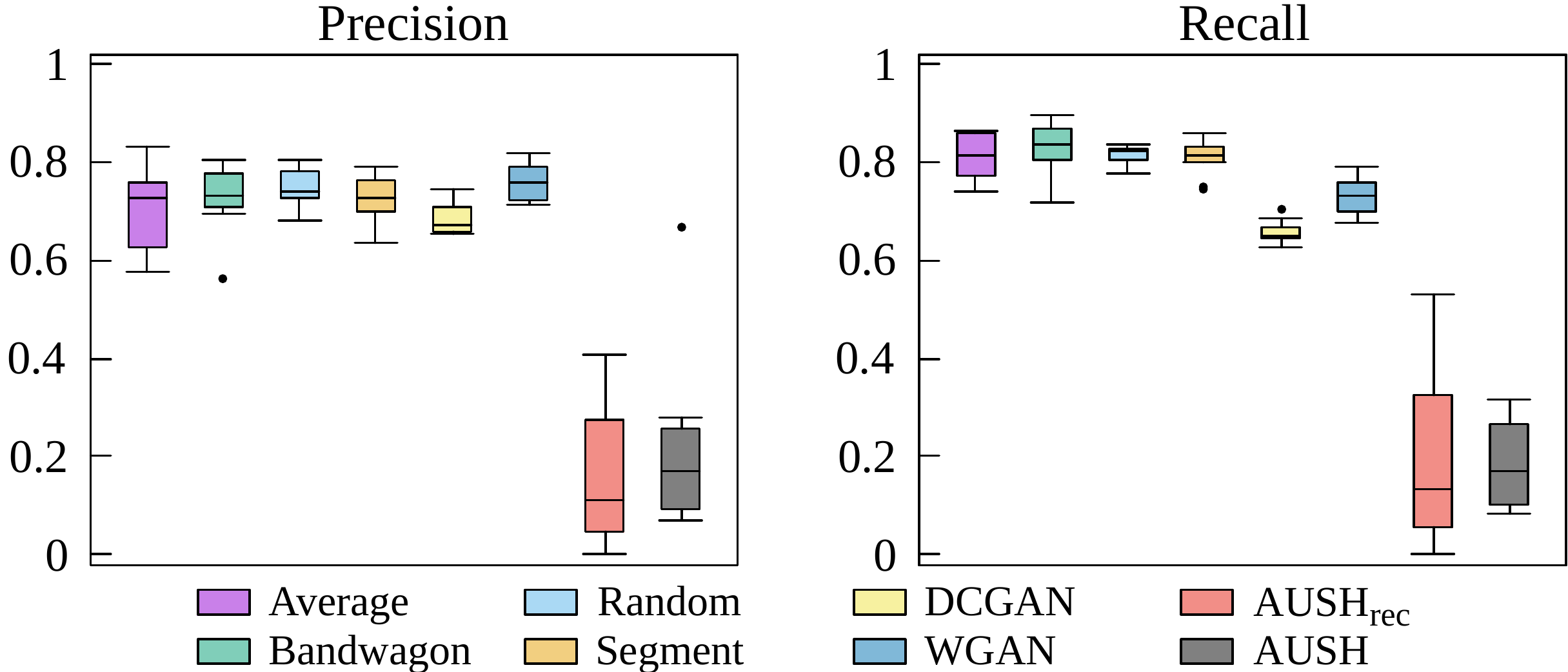}
\caption{Attack detection of injected profiles on ML-100K. Lower value suggests a better attack model.}
\label{fig:detectml}
\end{center}
\end{figure}

\subsection{Attack Detection (RQ5)}
\label{sec:detection}

% In order to testify the how realistic the injected user profiles can be,
We apply a state-of-the-art unsupervised attack
detector~\citep{Zhang2015Catch} on the injected user profiles generated by
different attack models and report the precision and recall on 10 random selected target
items. Fig.~\ref{fig:detectml} depicts the detection results on ML-100K. We can observe that the
detector performs the worst in terms of precision and recall against AUSH
and AUSH$_{rec}$, i.e., it fails to distinguish the injected user profiles generated by these two approaches.
On the contrary, most of the injected user profiles
from conventional attack models can be easily detected. Compared to AUSH, the detection performance of an ordinary
neural network such as AUSH$_{rec}$ is unstable over the 10 target items.
In the worst case, the injections generated by AUSH$_{rec}$ will be more
likely to be detected compared to those produced by AUSH. This observation further verifies the ability of our special designed AUSH to generate virtually undetectable injections in shilling attack.

%try table
%\begin{figure}[!t]
%\begin{center}
%\subfloat[TVD]{%
%\includegraphics[width=0.45\columnwidth]{ml100k_TVD}}
%\hspace*{\fill}
%\subfloat[JS divergence]{%
%\includegraphics[width=0.45\columnwidth]{ml100k_JS}}
%\caption{Distance between injected user profiles and real user profiles on ML-100K. }
%\label{fig:similarity}
%\end{center}
%\vspace{5pt}
%\end{figure}

Additionally, we run a set of similarity tests to further demonstrate the undetectability of AUSH.
We generate as many fake user profiles as the population of real users, i.e., 943 fake users for ML-100K. %We discretize the real ratings values into six bins $[-1.0, -0.6,-0.2,0.2,0.6,1.0]$ corresponding to the 5-point scale.
We compute the distribution $\mathbf{p}(v)\in\mathcal{R}^{6}$ for each item $v$, where $\mathbf{p}_i$ is the percentage of real ratings on $v$ with value $i, i=\{0,1,2,3,4,5\}$. We also compute the distribution $\mathbf{q}(v)$ in the injected user profiles.
Following~\citet{Christakopoulou2018Adversarial,Christakopoulou19}, we compute two distance measures, i.e., Total Variation Distance and Jensen-Shannon divergence:
\begin{small}
\begin{equation}
\begin{aligned}
TVD &= \sum_{v\in\mathcal{V}} \left|\mathbf{p}(v)-\mathbf{q}(v)\right|\big/\left|\mathcal{V}\right| \\
JS &=\sum_{v\in\mathcal{V}} \Big( \mathit{KL}\big(\mathbf{p}(v)\big\|\mathbf{m}(v)\big) + \mathit{KL}\big(\mathbf{q}(v)\big\|\mathbf{m}(v)\big) \Big)\big/|\mathcal{V}|\nonumber
\end{aligned}
\end{equation}
\end{small}
where $\mathbf{m}(v)=\big(\mathbf{p}(v)+\mathbf{q}(v)\big)/2$ and $\mathit{KL}(\cdot)$ represents the Kullback-Leibler divergence, between fake profiles and real profiles.
As shown in Tab.~\ref{tab:similarity}, the fake profiles generated by AUSH have the smallest TVD and JS.
% Aush$_{rec}$, which is the ordinary NN with only reconstruction loss, performs slightly worse than AUSH.
% This observation further verifies the ability of a GAN framework to generate virtually undetectable injections.
Since TVD and JS measure the difference of overall rating distributions, we can see that AUSH can preserve the distribution patterns and diversity of the original rating space.
\vspace{-2pt}

% \subsection{Impacts of Sampling Strategies and Losses}\label{sec:parameters}

% We also investigate the impacts of different sampling strategies (described in Sec.~\ref{sec:sampling}) and losses on the performance of attacks and report the results in Tab.~\ref{tab:result_sample} with best results highlighted in bold. For space reasons, we only show average results for attacking random targets and random long-tail targets in recommendation method I-AutoEncoder on ML-100K, but we observe similar results in other settings. AUSH$_{rand}$, AUSH$_{rating}$, AUSH$_{pop}$ and AUSH$_{sim}$ indicate random sample, sample by rating, sample by popularity and sample by similarity, respectively. In these methods, the complete loss $\mathcal{L}_{AUSH}$ is used. AUSH$_{basic}$ denotes that only adversarial loss is adopted and random sample is employed as the sampling strategy of filler items.

% We can observe from Tab.~\ref{tab:result_sample} that sample by rating is the
% best sampling strategy for AUSH and it produces the best attack results in
% both in-segment users and all users. The reason may be that it is easy to
% bias people with items having high ratings, as customers tend to trust such
% ``power'' items~\citep{SeminarioW14b}. On the other hand, AUSH$_{basic}$ has
% the worst attack performance compared to both other variations of AUSH and
% conventional attack approaches, showing that reconstruction loss and shilling
% loss indeed help improve the quality of generated profiles and increase the
% impact of shilling attack against RS.

\section{Conclusion}
\label{sec:con}

In this paper, we present a novel shilling attack framework AUSH. We design a
minimax game to let each of the attack profile generator and fake profile discriminator iteratively strikes to improve itself and beat the other one. We additionally employ a reconstruction loss and a shilling loss to help generate ``perfect'' fake profiles and achieve secondary attack goals. 
The experimental results show the superiority of
AUSH. 
In the future, we plan to design more sophisticated mechanisms
for learning selected items instead of selection by human. This way, the
ultimate goal of the attack can not be easily inferred from the selected items and
AUSH can become even more undetectable.

% \clearpage
\bibliographystyle{unsrtnat}
% Not sorted
% \bibliographystyle{ACM-Reference-Format}
% \setcitestyle{numbers,sort&compress}
% \balance

\bibliography{ref.bib}

\end{document}